\documentclass[fleqn,10pt]{wlscirep}
\usepackage[numbers,square]{natbib}
\usepackage{amsmath,amssymb,amsfonts,amsthm}
\usepackage{latexsym}
\usepackage{hyperref}
\usepackage{url}
\usepackage{xcolor}
\usepackage[utf8]{inputenc}
\usepackage[T1]{fontenc}
\usepackage{cleveref}
\usepackage{xr}
\usepackage{lineno}

\title{Instance-level quantitative saliency in multiple sclerosis lesion segmentation}

\author[1,2,3,4]{Federico Spagnolo}
\author[4,5,6]{Nataliia Molchanova}
\author[5,6]{Meritxell Bach Cuadra}
\author[1,2,3]{Mario Ocampo-Pineda}
\author[1,2,3]{Lester Melie-Garcia}
\author[1,2,3]{Cristina Granziera}
\author[4,+]{Vincent Andrearczyk}
\author[4,7,*,+]{Adrien Depeursinge}
\affil[1]{Translational Imaging in Neurology (ThINk) Basel, Department of Medicine and Biomedical Engineering, University Hospital Basel and University of Basel, Basel, Switzerland}
\affil[2]{Department of Neurology, University Hospital Basel, Basel, Switzerland}
\affil[3]{Research Center for Clinical Neuroimmunology and Neuroscience Basel (RC2NB), University Hospital Basel and University of Basel, Basel, Switzerland}
\affil[4]{MedGIFT, Institute of Informatics, School of Management, HES-SO Valais-Wallis University of Applied Sciences and Arts Western Switzerland, Sierre, Switzerland}
\affil[5]{CIBM Center for Biomedical Imaging, Lausanne, Switzerland}
\affil[6]{Radiology Department, Lausanne University Hospital (CHUV) and University of Lausanne, Lausanne, Switzerland}
\affil[7]{Nuclear Medicine and Molecular Imaging Department, Lausanne University Hospital (CHUV), Lausanne, Switzerland}

\affil[*]{adrien.depeursinge@hevs.ch}

\affil[+]{these authors contributed equally to this work}


\begin{abstract}
In recent years, explainable methods for artificial intelligence (XAI) have tried to reveal and describe models' decision mechanisms in the case of classification and even for segmentation. However, XAI methods for semantic segmentation and in particular for single specific instances (e.g. one given lesion among others of the same class in medical imaging) have yet to be developed to understand what drove the detection and contouring of the latter, which is crucial for all multi-lesional diseases.

We proposed instance-level explanation maps for semantic segmentation extending both SmoothGrad and Grad-CAM++ methods and yielding quantitative instance saliency for the former.
The instance-level methods were applied to the segmentation of white matter lesions (WML), a magnetic resonance imaging (MRI) biomarker in multiple sclerosis (MS).
687 patients diagnosed with MS for a total of 4023 FLAIR and MPRAGE MRI scans were collected at the University Hospital of Basel, Switzerland. 
WM lesion masks were annotated by four expert clinicians on baseline and follow-up imaging.
Three deep learning networks —a 3D U-Net, nnU-Net, and Swin UNETR— were trained and tested on these data (test normalized Dice score, respectively of 0.71, 0.78, 0.80; true positive rate of 79\%, 78\%, and 85\%; false discovery rate of 37\%, 38\%, and 36\%; false negative rate of 20\%, 22\%, and 14\%), then saliency maps were computed.

Consistent with clinical practice, the proposed instance saliency maps revealed that the model relied more on FLAIR than MPRAGE to segment WMLs, with positive saliency values inside a lesion and negative in its neighborhood. 
FLAIR hyperintensity combined with healthy WM around the lesion border was required for their detection.
Beyond the aforementioned sanity checks, we observed that peak values of the generated saliency maps presented distributions that differ significantly between TP, FN, FP and TN predictions, suggesting that the quantitative nature of the proposed saliency could be used to identify errors.

In conclusion, we introduced two XAI methods to generate quantitative instance-level explanations in semantic segmentation. 
The proposed XAI maps can be applied to any architecture and could serve as a basis to (i) improve model performance (e.g. reducing FPs), (ii) optimize their internal architecture (e.g. patch size), and (iii) justify the model's decisions to the end users, which are contextualized to a specific lesion instance of interest.
\end{abstract}

\begin{document}

\flushbottom
\maketitle
%
%
\thispagestyle{empty}


\section*{Introduction}


Multiple sclerosis (MS) is an autoimmune neurological disease, which affects people at a relatively young age, presenting a considerable impact on the quality of life~\cite{koltuniuk2023}. One of the most important biomarkers in MS are white matter (WM) lesions reported on magnetic resonance imaging (MRI)~\cite{yang2022}. The standard MRI sequences used for MS diagnosis and follow-up~\cite{thompson2017} are the fluid attenuated inversion recovery (FLAIR) and T1-weighted (T1-w) contrast, such as the magnetisation-prepared rapid gradient echo (MPRAGE)~\cite{hemond2018}. WM lesions generally appear hyperintense on FLAIR (except for cavitary lesions \cite{ayrignac2016}), and a subset with greater demyelination and tissue damage appears hypointense on T1-weighted images~\cite{thaler2015}.

\begin{figure}[!ht]
    \centering
    \includegraphics[width=0.5\textwidth]{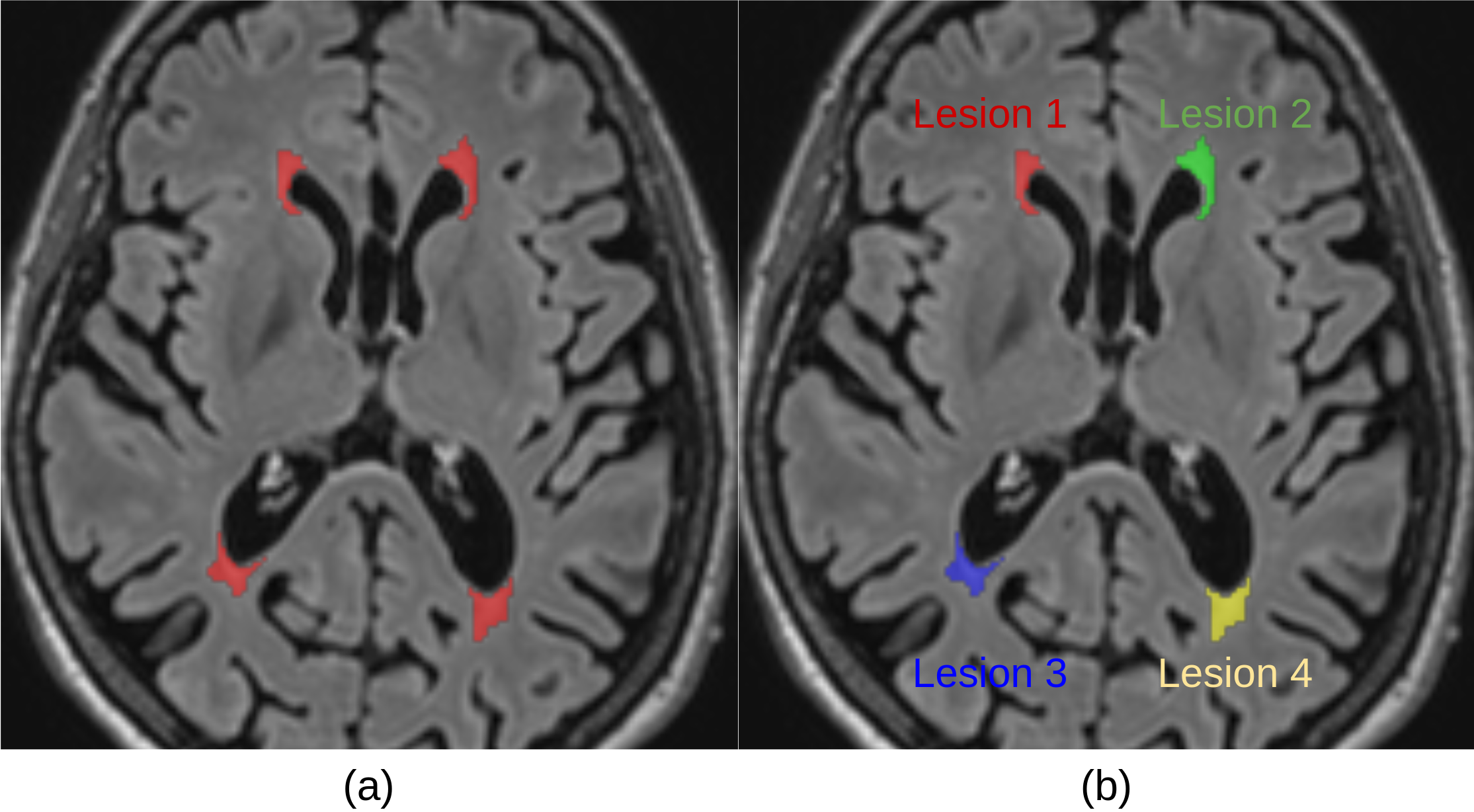}
    \caption{(a) FLAIR MRI presenting WM lesions segmented as separate entities, and (b) example of instance segmentation inspired by Varatharasan et al.\cite{varatharasan}.}
    \label{fig:instanceVSsemantic}
\end{figure}

These lesions are usually manually or semi-automatically annotated by clinicians with several years of experience, through a time-consuming process, subject to inter-observer variations. Despite many efforts to automate the process of lesion detection and segmentation with deep learning (DL) methods~\cite{review1,review2,review3,review4,review5}, their clinical integration is being jeopardized by two main issues:
\begin{enumerate}
    \item The ``black box'' nature of the models~\cite{baselli2020}. Since these methods contain many layers and millions of parameters it is hard to interpret and explain which are the drivers for a particular decision, i.e., which voxels were more important to identify and segment a given lesion of interest.
    \item Insufficient clinical validation of the models, described in Spagnolo et al.~\cite{spagnolo2023}.
\end{enumerate} 
To address the first issue, research in explainable AI (XAI) may play a decisive role. XAI has the potential to support trustworthy AI via better understanding (e.g., decision rules, biases) and optimization of DL models \cite{kobayashi2024}.
However, XAI for semantic segmentation and in particular for detecting and contouring single instances of interest has been little studied to date. Semantic segmentation is a computer vision task, where labels are associated with every pixel of an image. In the case of WM lesions, such as depicted in Fig.\ref{fig:instanceVSsemantic}a with four distinct lesion instances, the segmented plaques appear as disconnected volumes in the MRI. In general, objects may appear as connected, and can be segmented into separate entities through instance segmentation. Instance segmentation can extract supplementary information from the image, such as the number of objects of the same class (as lesion one to four in Fig.\ref{fig:instanceVSsemantic}b). 
Treating separate instances would be crucial to understand the mechanisms underlying automatic detection and segmentation of a given object of interest (e.g., a lesion in medical imaging or a person in a natural image) among other objects of the same class. This also applies to the case of MS lesions, where plaques can be subdivided based on their location in the brain, or the stage of the disease \cite{jonkman2015}. Specifically, considering separate instances would be important to generate instance-specific explanations not only for diagnosis at single time-points, but also for disease monitoring in follow-ups.

Introducing instance-level explanation methods could facilitate a better understanding of AI's segmentation of single instances. In addition to the case of MS, such methods could be applicable to any pathology involving sparse lesions, beyond MRI, and possibly to any segmentation task.

An exhaustive review of XAI models and applications can be found in Saranya et al.~\cite{saranya2023}. For convolutional neural networks (CNN) in a classification scenario, a widely used ad-hoc method is pixel attribution (or saliency maps), in which pixels are colored based on their contribution to the classification. To this end, vanilla gradients~\cite{simonyan2013} is a method based on forward and backward propagation through the network. First, an input image (for instance a 3D volume) is fed forward through the network and a class score is computed. Then, the gradient of the score with respect to the input --- or a layer --- is calculated backwards to form a map, which represents positive and negative contributions of input voxels to the classification of the image. Maps generated with this method are easy to compute and visualize, but also noisy and sensitive to small changes in the input~\cite{devries2023}. 
To partially address these problems, Smilkov et al.~\cite{smilkov2017} introduced a method called SmoothGrad (SG). 
A more stable output is obtained by feeding multiple noisy versions of the input image to the network and averaging the obtained saliency maps. Some recent applications can be found in Goh et al.~\cite{goh2021} and Agarwal et al.~\cite{agarwal2021}.

Another widely used XAI method for classification, is Grad-CAM~\cite{selvaraju2017}. The gradients of a class score --- with respect to activation maps of a given layer --- are spatially aggregated through global average pooling. This way, a weight is computed for each activation map, representing its relevance. These weights are then employed to calculate the visualization heatmap as a linear combination of the activation maps, followed by a ReLU to focus on features positively influencing the class score. However, this method presented a low accuracy when the image contained multiple instances of the same class. To overcome this, Chattopadhyay et al.~\cite{Chattopadhyay2018} described Grad-CAM++, where the weights are obtained through a weighted average of the gradients.

SG and Grad-CAM++ can be seen as complementary methods. The first provides local-level information and works well in identifying the impact of multiple input channels (or modalities), since the gradients flow all the way back to the inputs. Yet, as mentioned above, it may be too sensitive to changes in the input. The second can generate more stable heatmaps, and probe specific layers of the network. However, it also bears its disadvantages: 1) the choice of the intermediate layer can be not trivial, 2) the potential presence of early skip-connections in the architecture would be disregarded, 3) stopping the backpropagation at an intermediate layer would make it impossible to determine the impact of the input on the output. 

\begin{figure}[!ht]
    \centering
    \includegraphics[width=0.5\textwidth]{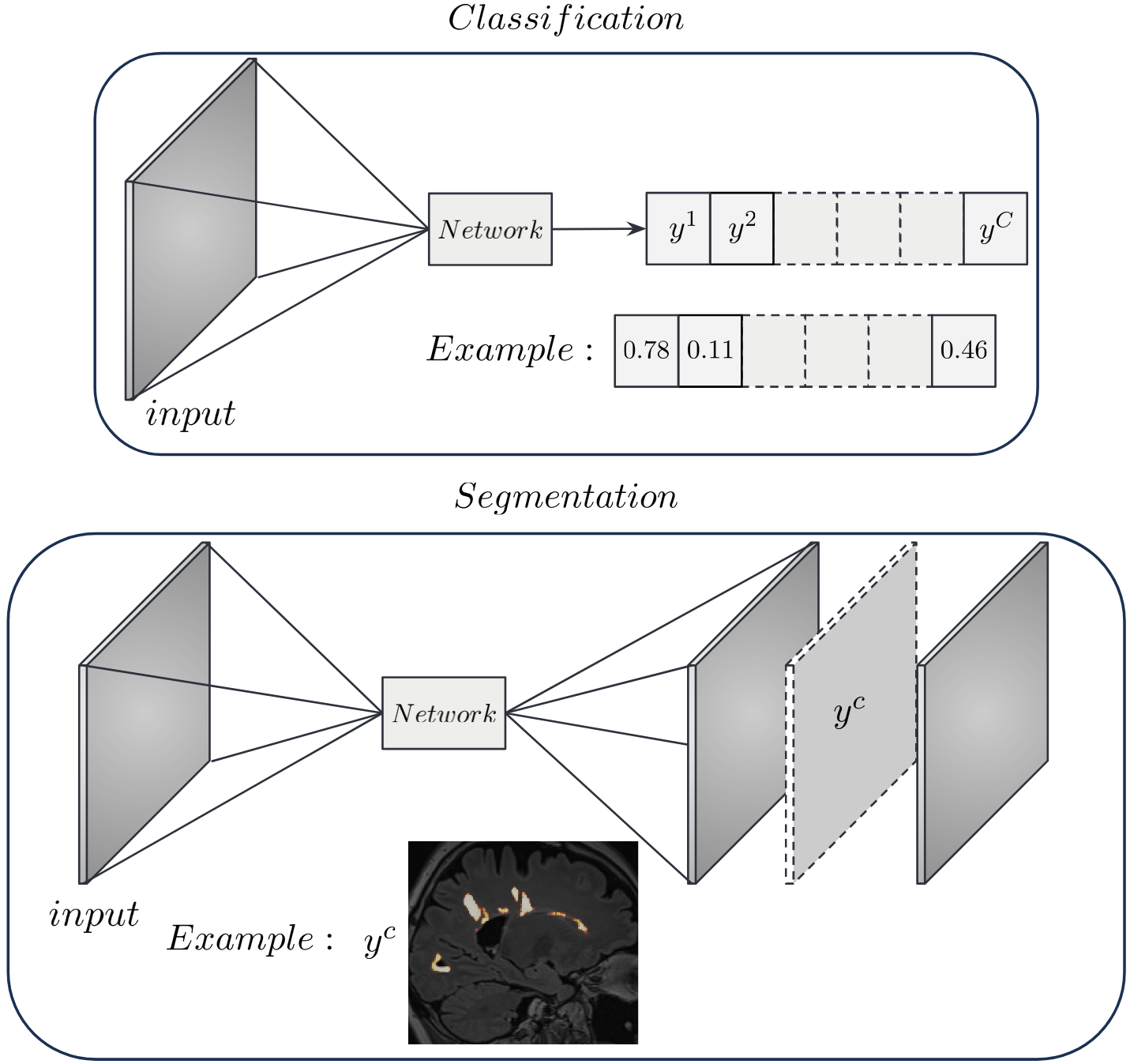}
    \caption{Input-output dimensions for a classification (top) and a semantic segmentation (bottom) task.}
    \label{fig:classvssegm}
\end{figure}

Both algorithms were originally designed for classification tasks, such as in Rajpurkar et al.~\cite{rajpurkar2017}. This means that, in general, the output $y$ of a multi-class classification network is a vector of $C$ scalar values, where $C$ is the number of  classes. In the case of semantic segmentation, $y$ is a set of $C$ tensors (2D or 3D images) containing segmentation scores for each class (Fig.~\ref{fig:classvssegm}).

Indeed, computing a gradient map for all the output voxels would not be convenient or meaningful in a segmentation scenario. The number of maps per patient would be excessive, thus impractical for the end-user, especially for a clinician. Additionally, the computation time would be unnecessarily high. 

\begin{figure}[!ht]
    \centering
    \includegraphics[width=0.5\textwidth]{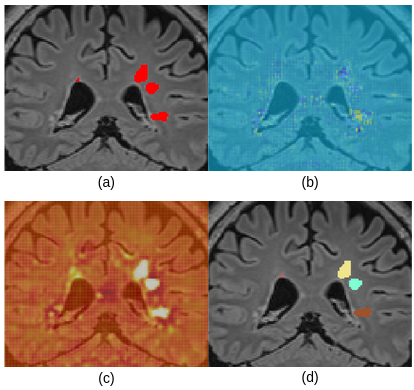}
    \caption{(a) The output of a semantic segmentation network showing several instances of the considered class. SmoothGrad (b) and Grad-CAM (c) applied to all the spatial predictions. (d) How can we explain the  segmentation of a particular lesion of interest (e.g., the yellow instance)?}
    \label{fig:introgen}
\end{figure}

A straightforward way to adapt explainable methods to semantic segmentation would be, for a given class $c$, to aggregate all the spatial predictions $y^{c}$ into a single scalar (e.g., a summation) and compute its gradients. However, this approach yields confusing and hardly interpretable maps, where the relevance of input voxels to the segmentation scores of all output voxels (even those not segmented as part of class $c$) are merged together, as reported in Fig.~\ref{fig:introgen}. This makes it more complicated to understand and determine which part of the output segmentation is influenced by which region of the input. That is especially important when there are multiple objects belonging to the same class, so that input intensity of neighboring voxels or objects may impact the segmentation. Indeed, these results do not provide any kind of explanation to the segmentation of a particular instance of the considered class $c$. 

Recent works~\cite{mahapatra2022,arun2021} clearly state that saliency maps were not initially developed for segmentation tasks and, therefore, no explicit methods are currently available to do so. In Arun et al.~\cite{arun2021}, this statement was used to warn the readers about potential problems related to the misuse of XAI in clinical practice, and proposed pixel-level metrics to evaluate saliency maps. Singh et al.~\cite{singh2022} applied saliency maps to classification networks, while adopting uncertainty maps to assess the confidence of their skin lesion segmentation method.

First steps towards the use of XAI in segmentation were taken in the work of Wickstrom et al.~\cite{wickstrom2020}. They used guided backpropagation~\cite{simonyan2013} to generate saliency maps for the explanation of colorectal polyp segmentation. Their saliency maps were obtained by aggregating all the positive spatial predictions. Similarly, Vinogradova et al.~\cite{vinogradova2020} generated heatmaps by using the first version of Grad-CAM~\cite{selvaraju2017}, and aggregating only output voxels segmented as part of a target class~$c$. The authors employed a U-Net and the dataset Cityscapes~\cite{cordts2016} to perform semantic segmentation. 

While these approaches lead to a possible class-level explanation for semantic segmentation, they still do not provide any instance-level information, such as which input voxels were exploited to segment a specific instance. In addition, these methods do not provide quantitative saliency maps, which means these maps do not allow to interpret their absolute values across images or objects of interest to, e.g., distinguish false positives or false negatives from true positives.

To address the aforementioned limitations, this paper presents the adaptation of SG and Grad-CAM++ to obtain quantitative and instance-level explanation maps, and their application to WM lesion segmentation in MS. 

The quantitative characteristic of the saliency method described in Section~\hyperlink{max}{Quantitative saliency maps: maximum versus average aggregation} was exploited in Spagnolo et al.~\cite{spagnolo2025}, using the same MRI data. Radiomic features extracted from the proposed instance-level XAI maps were fed to a simple logistic regression model to refine the classification of TP and FP examples. In that study, a bootstrapping approach with 1000 iterations was used to compute the 95\% confidence intervals, highlighting an F1 score relative improvement on the test set.

\section*{Methods}


\subsection*{Dataset and models}

687 patients diagnosed with MS for a total of 4023 FLAIR and MPRAGE MRI scans (age=45.2±12.2, 433 females, SIEMENS Avanto/Espree/Symphony 1.5T and Prisma/Skyra/Verio/MAGNETOM Vida 3T, $1mm$ isotropic, Expanded Disability Status Scale median of 2.5 [0-9]) were collected at the University Hospital of Basel, Switzerland~\cite{SMSC}. The study received ethical approval by the local independent ethics committee (Ethikkommission Nordwest- und Zentralschweiz, EKNZ), all patients provided written informed consent, and all methods were carried out by relevant guidelines and regulations.

The image size of both MR contrasts is (192 × 240 × 256), which corresponds to a volume of (192 × 240 × 256) $mm^3$. Table \ref{supptable1} in the Supplementary Material reports data information for each MR system, such as age, sex and number of visits. For consistency reasons, patients were mainly scanned first using Avanto 1.5T and then Skyra 3T, operating other MR systems only in case of unavailability of the two mentioned. 

WM lesion masks from baseline and follow-ups were semi-automatically annotated by four expert clinicians (>5 years of experience), independently and without consensus. The binary lesion masks corrected by the experts were generated by a U-Net variant described in La Rosa et al. (2020)\cite{larosa2020} (since that variant was trained on FLAIR and MP2RAGE data, we retrained it on separate proprietary FLAIR and MPRAGE data before the inference).
Data were randomly split into training, validation and test sets (containing 560, 90 and 37 patients with 3369, 553 and 101 scans, respectively; training, validation and hold-out test set’s mean lesions number of $52.0\pm36.3$, $56.2\pm35.3$, and $42.3\pm21.4$ per patient) to train and evaluate a 3D U-Net~\cite{cicek2016} variant (different from the one in La Rosa et al \cite{larosa2020}), an nnU-Net \cite{isensee2021}, and a Swin UNETR \cite{hatamizadeh2022} for WM lesion segmentation, using patches of dimensions $96^{3}$ to ensure the inclusion of at least part of the brain in every patch. The split was performed at patient level to ensure that images from the same patient belong to the same split. We adopted a linear combination of normalized Dice~\cite{raina2023} and blob~\cite{kofler2022} losses to tackle instance imbalance within a class and bias towards the occurrence of positive class~\cite{maier-hein2022}.  
Pre-processing steps included the registration of FLAIR images to MPRAGE space using the \textit{elastix} toolbox~\cite{klein2009,shamonin2014}, N4 bias field inhomogeneity correction~\cite{tustison2010} and $z$-score intensity normalisation.

\subsection*{Notations}

Following Depeursinge et al. (2020)~\cite{depeursinge2020}, 
we noted a discrete image as a $D$-dimensional function of the variable \mbox{$\boldsymbol{v} = (v_1, \dots , v_D) \in \mathbb{Z}^D$}, taking values $x[\boldsymbol{v}] \in \mathbb{R}$. 
A subset $\Gamma$ of $\mathbb{Z}^D$ was considered in practice for the spatial image domain with dimensions $N_1\times\cdots\times N_D$ as possible values for the index vector $\boldsymbol{v}\in\Gamma$. We also referred to the lesion domain $\Omega$ as a subset of the image domain with cardinality (i.e. number of voxels) $|\Omega |$, such that $\Omega\subset\Gamma\subset\mathbb{Z}^D$.

Input images $x[\boldsymbol{v}]$ performing a forward pass through the network resulted in logits $y\left(x\right) [\boldsymbol{v}] \in \mathbb{R}$, where $\boldsymbol{v}\in\Gamma$ is a map of raw output values, which had the exact same dimensions as the input values $x[\boldsymbol{v}]$ since we considered a segmentation task. We use the simplified notation $y[\boldsymbol{v}]$ when the input $x$ is unambiguous.
This raw output was generally interpreted as a probability map (e.g., after Softmax), which was binarized using a threshold $t=0.3$ yielding the best normalized Dice score during validation. Various thresholds $t$ were explored, ranging from 0.1 to 0.5, and the one selected scored about 2\% better than the second best. Then, each connected component of the binary map was considered as a specific WM lesion instance, forming $\Omega$.

\subsection*{Gradient-based saliency maps} 

During the backward pass and for each voxel of $y\left(x\right)[\boldsymbol{v}]$, the gradients with respect to all the voxels of input values $x[\boldsymbol{v}]$ can be computed. These gradients can be visualized as an image with same dimensions as the input and output, constituting a first method to construct saliency maps. This approach is usually referred to as vanilla gradients~\cite{simonyan2013}.

The SG algorithm~\cite{smilkov2017} tackled the problem of instability of vanilla gradients maps: the gradient of logits $y\left(x_n\right)$ is computed $N$ times based on artificially noised versions of the input $x_n[\boldsymbol{v}]$. The authors demonstrated that the average $M$ of these maps
is more stable than vanilla gradients maps:

\begin{equation} \label{smoothgrad}
    M [\boldsymbol{v}] = \frac{1}{N} \sum_{n=1}^{N} \frac{\partial y (x_{n})}{\partial x_{n}[\boldsymbol{v}]}.
\end{equation}

\subsubsection*{Instance-level saliency (gradients)}

\begin{figure}[!ht]
    \centering
    \includegraphics[width=0.7\textwidth]{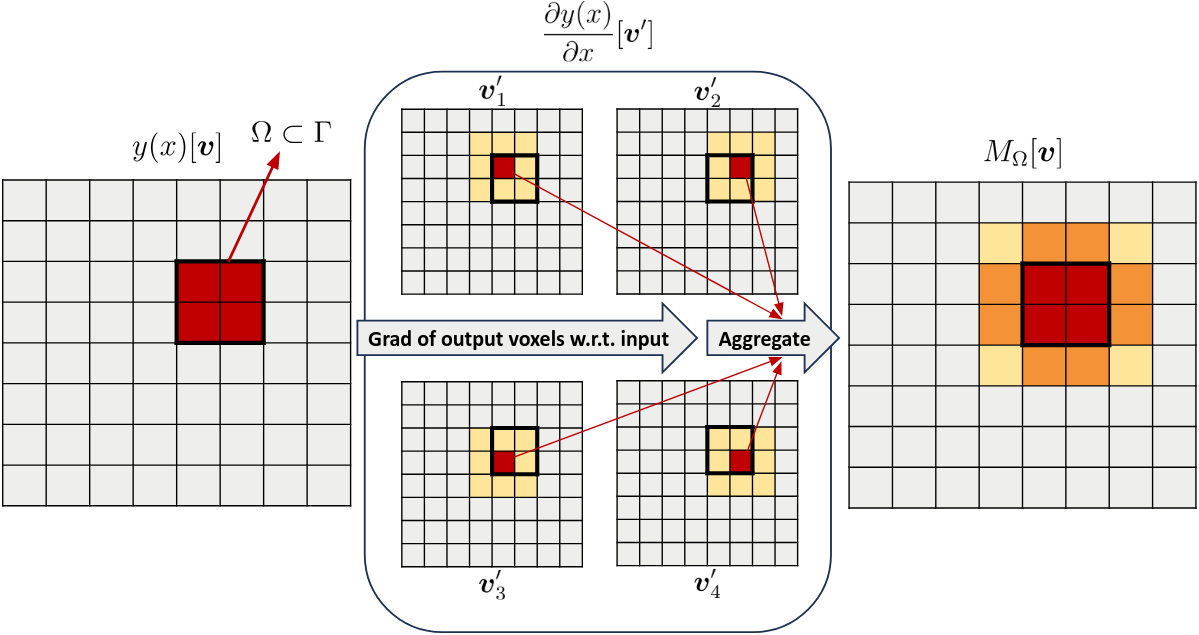}
    \caption{Overview of the proposed adaptation of SG to segmentation.}
    \label{fig:smooth_flowchart}
\end{figure}

This approach was introduced in a classification paradigm. However, with segmentation models, there are predictions for each output voxel $\boldsymbol{v}$. As a result, the visualisation of many heatmaps for a single output voxel is neither convenient nor meaningful. Hence, we adapted the original method to the segmentation task by aggregating these heatmaps. For a given lesion, the implementation consists of:
\begin{enumerate}
    \item Injecting a Gaussian noise $\mathcal{N}(0,\sigma)$ with standard deviation $\sigma$ to obtain a noisy version of the input,
    \item Computing a collection of saliency maps for all output voxels in the domain $\Omega$ of the lesion,
    \item Determining the average map from this collection of maps,
    \item Repeating steps 1-3 and combining $N=50$ saliency maps ($\sigma=0.05$) from $N$ noisy versions to obtain a single one.
\end{enumerate}
Steps 2 and 3 are illustrated in Fig.~\ref{fig:smooth_flowchart}. Eq.\eqref{localsmoothgrad} details the computation of lesion-level saliency maps $M_{\Omega}^{\text{gradient}}[\boldsymbol{v}]\in\mathbb{R}$ combining gradients calculated from each output voxel of a lesion. A separate saliency map is generated for each input modality (two in our case) allowing us to investigate their respective contribution.

\begin{equation} \label{localsmoothgrad}
    M_{\Omega}^{\text{gradient}}[\boldsymbol{v}] = \frac{1}{N|\Omega|} \sum_{n=1}^{N} \sum_{\boldsymbol{v}'\in\Omega} \frac{\partial y(x_{n})[\boldsymbol{v}']}{\partial x_{n}[\boldsymbol{v}]}
\end{equation}

\subsubsection*{Quantitative saliency maps: maximum versus average aggregation}
\hypertarget{max}{}

The advantage of $M_{\Omega}^{\text{gradient}}[\boldsymbol{v}]$ is that it shows whether voxels outside the lesion domain $\Omega$ impact the prediction of voxels belonging to $\Omega$. However, the lesion domain's dimensions may range from a few voxels to a considerable volume. The greater the lesion size, the more extended the potential distance between two voxels $p,q \in \Omega$. A long-distance means that the saliency map generated for the voxel $p$ will present a low gradient value for $q$. The same principle applies to regions of the input far away from a lesion: their contribution to the prediction is low. As a consequence, the average over the lesion domain $\Omega$ in Eq.\eqref{localsmoothgrad} will cause gradient values in $M_{\Omega}^{\text{gradient}}[\boldsymbol{v}]$ for extensive lesions to be systematically lower. Thus, following this method, the voxel values in saliency maps generated for different $\Omega$ (lesions) would not be comparable.

To this end, we propose Eq.\eqref{maxsmoothgrad}, a slightly modified version of Eq.\eqref{localsmoothgrad} where the average of saliency maps generated from each element of $\Omega$ is replaced by the voxel-wise maximum with sign:

\begin{equation} \label{maxsmoothgrad}
    M_{\Omega}^{\text{gradient}}[\boldsymbol{v}] = \frac{1}{N} \sum_{n=1}^{N}D^{n}_{argmax_{\boldsymbol{v}'}|D^{n}_{\boldsymbol{v}'}|}, \text{ where } D^{n}_{\boldsymbol{v}'} = \frac{\partial y(x_{n})[\boldsymbol{v}']}{\partial x_{n}[\boldsymbol{v}]}
\end{equation}

\subsection*{Saliency maps based on Grad-CAM++}

\begin{figure}[!ht]
    \centering
    \includegraphics[width=0.8\textwidth]{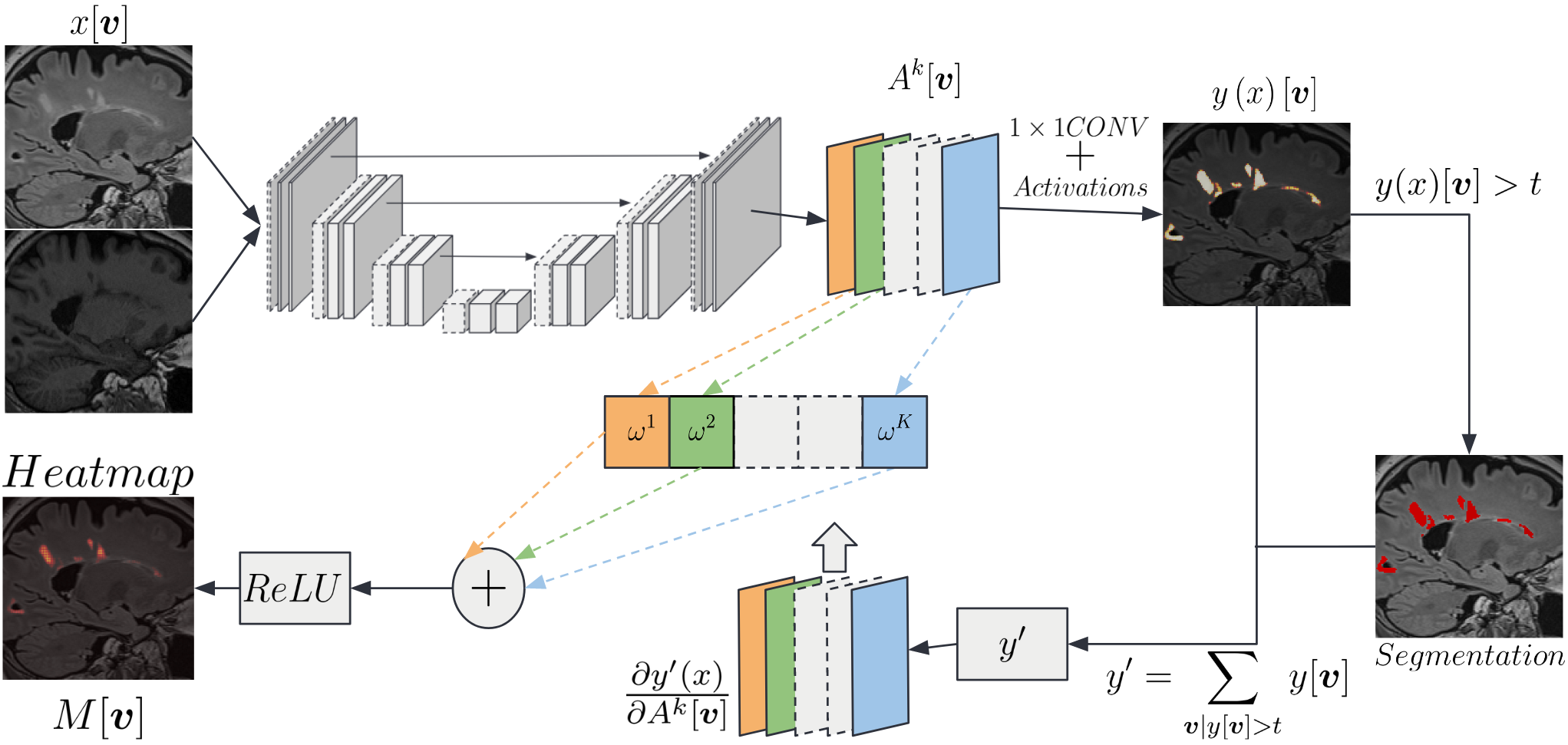}
    \caption{Overview of Grad-CAM++, generating a class-level explanation heatmap, similarly to Vinogradova et al.~\cite{vinogradova2020}, adapted for segmentation.}
    \label{fig:orig_gradcam_flowchart}
\end{figure}

The second proposed method is based on Grad-CAM++~\cite{Chattopadhyay2018}. We generated, for a given layer of a network, a heatmap $M$ as a linear combination between weights $\{\omega^{k}\}^{K}_{k=1}$ and activation maps $\{A^{k}\}^{K}_{k=1}$. The activation maps of the selected layer may have different dimensions. In this case, the final heatmap is upsampled to the input dimensions. 
For the sake of simplicity, we defined the domain of the activation maps and that of the input image to be the same (i.e. activations from the last layer since we have a segmentation architecture): $\Gamma$, with values $A^{K}[\boldsymbol{v}] \in \mathbb{R}$. Following Vinogradova et al.~\cite{vinogradova2020}, to compute the gradients, we considered $y'$ as the sum of logits $y[\boldsymbol{v}]$ higher than a threshold $t$, as in Eq.\eqref{y-original}. Then, each weight $\omega^{k}$ is computed using the gradients of $y'$ - with respect to the $k^{th}$ activation map $A^{k}$ - and a coefficient $\alpha^{k}[\boldsymbol{v}]$. We have

\begin{equation} \label{map-original}
    M^{\text{GradCAM}}[\boldsymbol{v}] = Relu\left(\sum_{k}\omega^{k} \cdot A^{k}[\boldsymbol{v}]\right), 
\end{equation}
\begin{equation} \label{y-original}
    y' = \sum_{\boldsymbol{v}|y[\boldsymbol{v}]>t} y [\boldsymbol{v}],
\end{equation}
\begin{equation} \label{omega-original}
    \omega^{k} = \sum_{\boldsymbol{v}\in\Gamma}\alpha^{k} [\boldsymbol{v}]\cdot Relu\left(\frac{\partial y'}{\partial A^{k}[\boldsymbol{v}]}\right),
\end{equation}
\begin{equation} \label{alpha-original}
    \alpha^{k} [\boldsymbol{v}] = \frac{\frac{\partial^{2} y'}{\partial \left(A^{k}[\boldsymbol{v}]\right)^{2}}}{2 \cdot \frac{\partial^{2} y'}{\partial \left(A^{k}[\boldsymbol{v}]\right)^{2}} + \sum_{\boldsymbol{v}'\in\Gamma}\left(A^{k}[\boldsymbol{v}'] \cdot \frac{\partial^{3} y'}{\partial \left(A^{k}[\boldsymbol{v}]\right)^{3}}\right)},
\end{equation}
where $\boldsymbol{v}$ and $\boldsymbol{v}'$ correspond to a different indexing over the domain $\Gamma$. The above \cref{map-original,y-original,omega-original,alpha-original} are derived from Chattopadhyay et al. (2018)~\cite{Chattopadhyay2018}. A graphical representation of this method is reported in Fig.~\ref{fig:orig_gradcam_flowchart}.

\subsubsection*{Instance-level saliency (Grad-CAM++)}

\begin{figure}[!ht]
    \centering
    \includegraphics[width=0.8\textwidth]{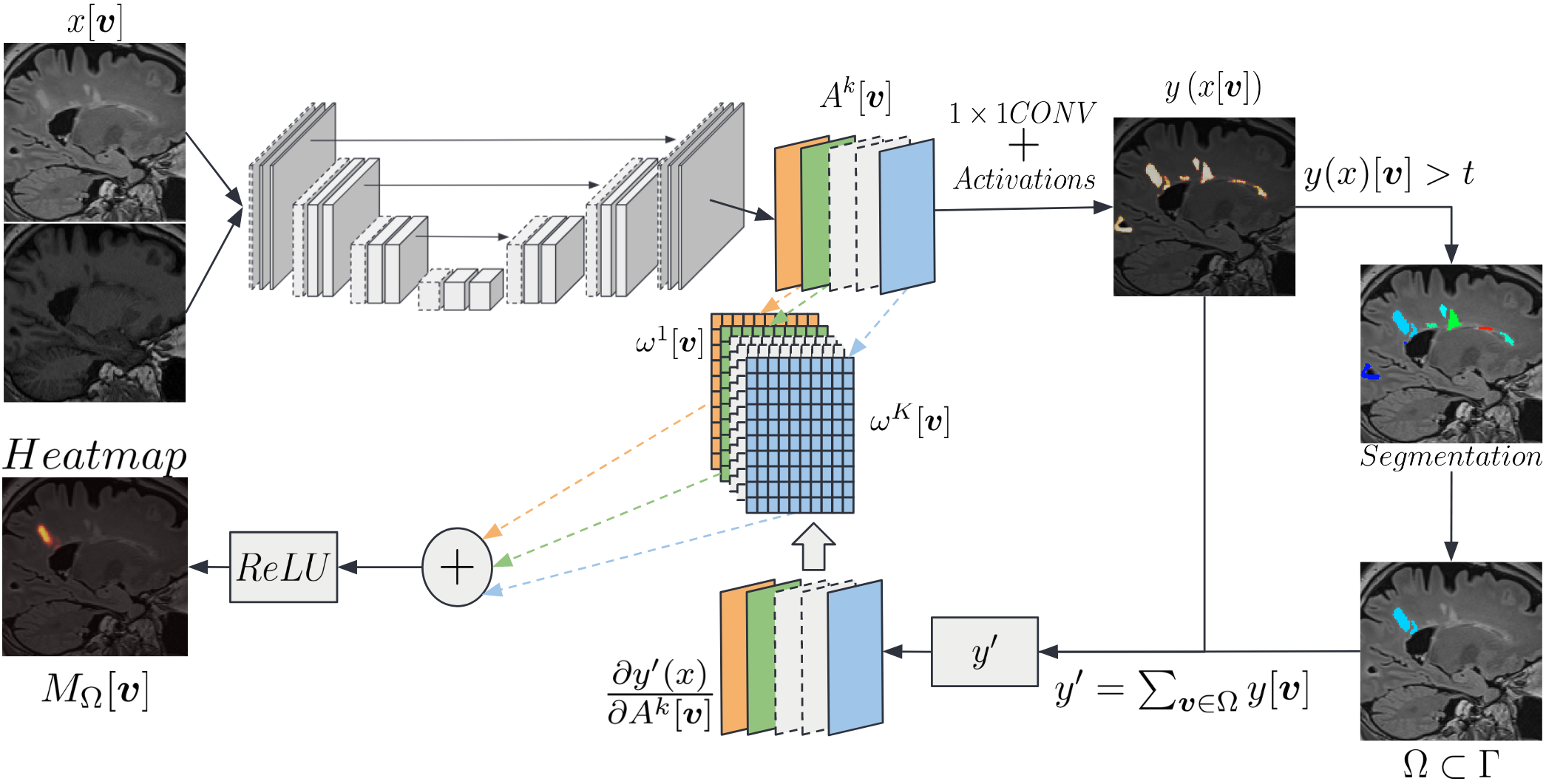}
    \caption{Overview of the proposed adaptation of Grad-CAM++, generating an instance-level explanation heatmap.}
    \label{fig:local_gradcam_flowchart}
\end{figure}

The simple adaptation of Grad-CAM++ to segmentation presented above generated a class-level explanation for semantic segmentation by merging contributions to different lesion instances. The impact of input regions on different parts of the output was combined by: 
\begin{itemize}
    \item Considering the gradients of a subset $y'$ of logits $y[\boldsymbol{v}]$,
    \item Assigning a single weight $\omega^{k}$ for each feature map $A^{k}$.
\end{itemize}
However, it would be useful to know which input voxels influenced the segmentation of a given instance (e.g., a lesion).
To adapt the algorithm to an instance-level explanation, we considered two steps. First, the gradients of $y$ were computed from the domain $\Omega$ of one lesion, as in Eq.\eqref{y-local}. Then, the summation in Eq.\eqref{omega-original} over $\Gamma$ to compute weights was removed to retain a weight for each element of a feature map. This was needed to prevent the activation of other instances to emerge, and to select only the activation in $\Omega$. Eq.\eqref{map-local} and Eq.\eqref{omega-local} represent the proposed heatmap $M_l^{\text{GradCAM}}$ provided by the modified Grad-CAM++ method, and the weights $\omega^{k} [\boldsymbol{v}]$:

\begin{equation} \label{map-local}
M_{\Omega}^{\text{GradCAM}}[\boldsymbol{v}] = Relu\left(\sum_{k}\omega^{k}[\boldsymbol{v}] \cdot A^{k}[\boldsymbol{v}]\right),
\end{equation}
\begin{equation} \label{y-local}
    y' = \sum_{\boldsymbol{v}\in\Omega} y [\boldsymbol{v}],
\end{equation}
\begin{equation} \label{omega-local}
    \omega^{k}[\boldsymbol{v}] = \alpha^{k}[\boldsymbol{v}] \cdot Relu\left( \frac{ \partial y'}{\partial A^{k}[\boldsymbol{v}]}\right),
\end{equation}

where $\alpha^{k}[\boldsymbol{v}]$ were obtained as in Eq.~\eqref{alpha-original}.
An overview of the proposed adaptation of Grad-CAM++ to segmentation is illustrated in Fig.~\ref{fig:local_gradcam_flowchart}. Comparing Figs~\ref{fig:orig_gradcam_flowchart} and~\ref{fig:local_gradcam_flowchart}, it can be observed that the output map of the first is a valuable explanation for the entire lesion class, combining the attribution of voxels from different lesions in the same map. Conversely, the output of the second separates the attribution of voxels from single lesions to obtain a more meaningful map, in cases where each instance should be treated independently.

The code for the computation of explainable maps through both methods is publicly available at the following GitHub repository: \url{https://github.com/federicospagnolo/IES.git}.

\subsection*{Experiments}

The described methodologies were applied to all the three selected architectures (U-Net, nnU-Net and Swin UNETR), obtaining a collection of saliency maps. The XAI maps presented below were computed using the U-Net for a total of 3050 true positive (TP), 1818 false positive (FP), 789 false negative (FN), and 1010 true negative volumes (TN). Examples of saliency maps derived from nnU-Net and Swin UNETR are included in the Supplementary Material. TP and FP predictions were defined as having, respectively, a non-zero and zero overlap with ground truth (GT). FN predictions were considered as GT segmentations with zero overlap with the predicted lesion mask. TN examples were generated by randomly sampling ten spherical volumes ($93 mm^{3}$) inside each patient's brain and skull, excluding volumes intersecting GT and prediction masks. The size of the TN volumes was decided based on the average lesion volume in the GT masks of the test set. In this study, when mentioning TP, FP, FN and TN we refer to examples, i.e. volumes, not voxels. While TN voxels represent most of the image, we tackle class imbalance by undersampling, randomly selecting ten spherical volumes per patient.

As for the minimum considered lesion size, the McDonald's criteria recommend a minimum in-plane diameter of 3 $mm$ based on old 2D sequences~\cite{polman2005,grahl2019}, and in literature there is no clear consensus. As a trade off between limiting partial volume effects and the number of false positives~\cite{fartaria2018}, a minimum volume size of $5 mm^{3}$ was set for each connected component, using a connectivity of 18. Several publications have selected a similar minimum lesion volume, for example: 1) De Rosa et al. (2024)\cite{derosa2024} considered 0.005 $ml$ (5 $mm^3$); 2) Fartaria et al. (2019)\cite{fartaria2018} used 0.006 $ml$; 3) Jain et al. (2016)\cite{jain2016} removed candidates with a volume smaller than 0.005 $ml$. 

The information exploited by the U-Net during inference for the segmentation of specific WM lesions was assessed with several experiments. Throughout these tests, the max aggregation method (see Section \hyperlink{max}{Quantitative saliency maps: maximum versus average aggregation}) was selected to be used, due to its ability to provide quantitative information. This only applies to the saliency based on SG. 

First (results in Section \hyperlink{Assessing the contributions of input MR sequences using gradient-based saliency maps}{Assessing the contributions of input MR sequences using gradient-based saliency maps}), the distribution of positive and negative values in saliency maps was observed to reveal potential patterns concerning recurring locations of positive/negative values with respect to $\Omega$. An analysis of these distributions allowed a statistical comparison between gradient values computed with respect to FLAIR and MPRAGE. As a consequence, it was possible to isolate the contribution of both input sequences to the prediction of single lesions. In this analysis, we discarded gradient values between -0.1 and 0.1 in order to focus on voxels with a higher attention level.

Second (results in Section \hyperlink{Statistical distribution of gradient values for TPs, FPs, FNs and TNs}{Statistical distribution of gradient values for TPs, FPs, FNs and TNs}), we quantitatively compared the distribution of maximum and minimum saliency maps' values, for all predictions categories (i.e. TP, FP, FN and TN) to investigate if absolute saliency values can be used to flag potential detection errors. Furthermore, a two-sided Mann Whitney U test~\cite{mcknight2010} was run to statistically compare these groups.

Third (results in Section \hyperlink{Sanity checks}{Sanity checks}), we tested the saliency method's behavior in two cases: (a) positioning a domain $\Omega$ in a region presenting no MS lesions (i.e. healthy WM); (b) considering a domain $\Omega$ of a single voxel, located at the center of mass of a true lesion. For both cases, we show the generated instance-level saliency and its range of values.

To understand the U-Net's level of specificity in segmenting MS lesions in the WM, we designed three additional qualitative experiments on a batch of 10 patients, in which we observed the U-Net's behavior on synthetic lesions (results in Section \hyperlink{Sanity checks}{Sanity checks}). A first experiment was conducted, moving a clearly visible WM lesion to a different part of the WM, which originally presented no lesions. A similar approach was followed when inserting the same lesion outside the skull. A third experiment consisted of a compromise: the lesion, along with part of its surrounding healthy tissue ($3 mm$ from lesion border), was moved outside the skull. For all these cases, the U-Net's prediction and saliency maps were examined.

In light of the results observed during the described tests, we designed a more specific experiment on the entire test set, using all the three trained networks: the analysis of the needed amount of contextual information surrounding a lesion to obtain a segmentation (results in Section \hyperlink{Experiment on contextual information}{Experiment on contextual information}). We selected lesions with size close to the average of the entire dataset, i.e. from $90$ to $120 mm^{3}$, obtaining a total of 191 (U-Net), 173 (nnU-Net), and 161 (Swin UNETR) TP lesions. Initially, all voxel intensities were set to zero, but those of the lesion: this step was called iteration 0. Then, we gradually reassigned the original intensity to surrounding voxels through morphological 3D dilation, iterating this process 35 times. With the final iteration, the models were seeing a volume of surrounding tissue at a maximum of $35 mm$ distance from the lesion's edges. At each iteration, the following metrics were recorded: 
\begin{itemize}
    \item The average and standard deviation across lesions of the mean prediction score (after Softmax) in $\Omega$,
    \item The number of segmented lesions.
\end{itemize}
In this experiment, each lesion was considered detected when at least one voxel in $\Omega$ reached a prediction score above the threshold $t=0.3$.



\section*{Results}

The trained lesion segmentation models —U-Net, nnU-Net, Swin UNETR— achieved, respectively, a test Dice score of 0.60, 0.62, 0.66, a normalized Dice score of 0.71, 0.78, 0.80, a true positive rate of 79\%, 78\%, and 85\%, a false discovery rate of 37\%, 38\%, and 36\%, a false negative rate of 20\%, 22\%, and 14\%. The count of TP, FP and FN examples was 3050, 1818, 789 for the U-Net, 3112, 1954, 880 for the nnU-Net, and 3516, 2018, 598 for the Swin UNETR.

\subsection*{Maximum versus average aggregation for quantitative saliency maps}
\hypertarget{Maximum versus average aggregation for quantitative saliency maps}{}

\begin{figure}[!ht]
    \centering
    \includegraphics[width=0.5\textwidth]{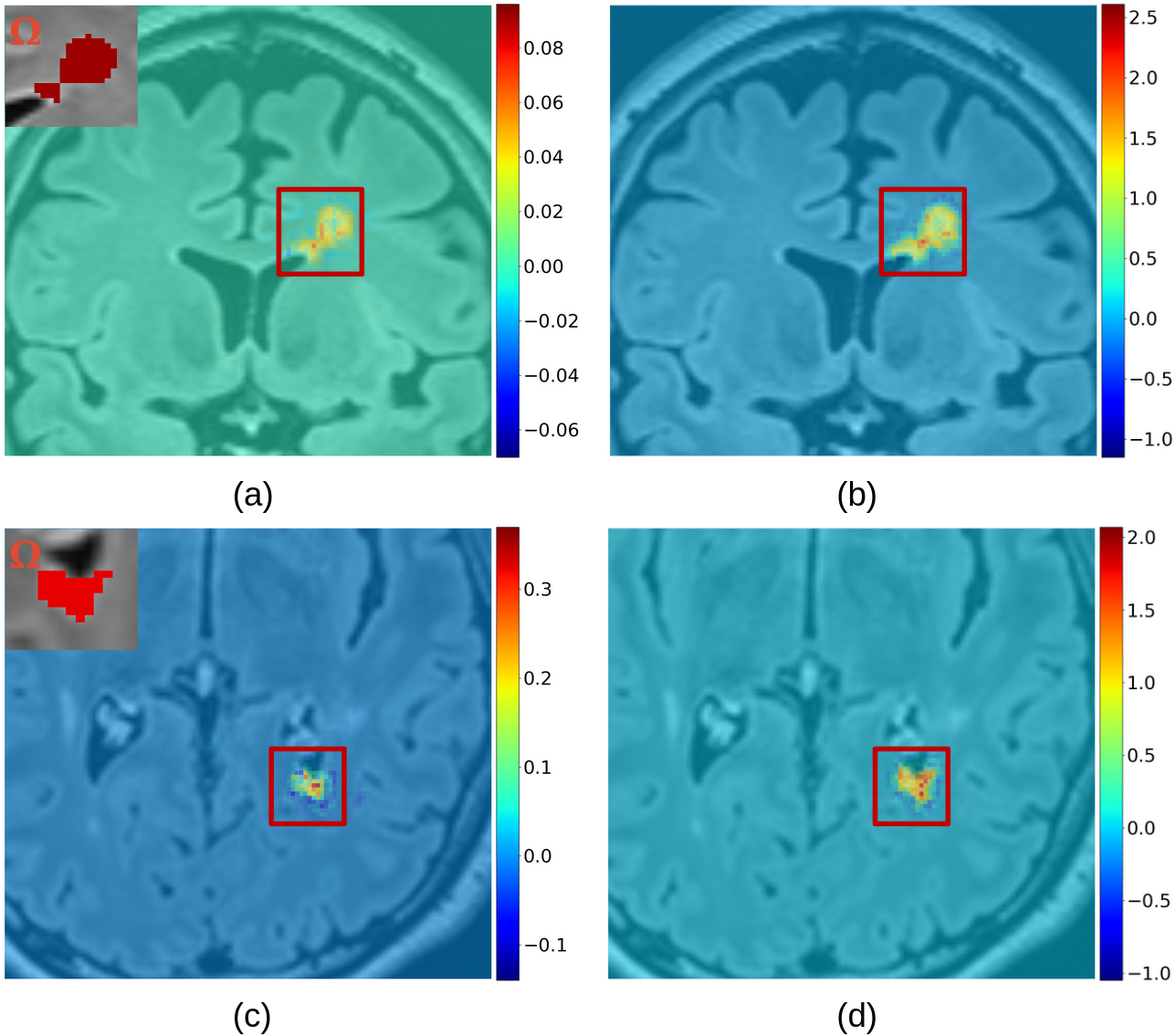}
    \caption{Saliency maps (computed for U-Net) based on SG generated for an extensive lesion, following the average (a) and the maximum (b) methodology. Saliency maps generated for a smaller lesion, following the average (c) and the maximum (d) methodology. The gradients intensity difference between (a,c) and (b,d) can be observed from the color scales range next to each map, increasing approximately from (-0.1, 0.3) to (-1, 2.5). The thumbnails illustrate the lesion domains $\Omega$ used during the computation.}
    \label{fig:max-vs-avg}
\end{figure}

Considering only saliency based on SG, we compared the saliency maps obtained with either the average (Eq.\eqref{localsmoothgrad}) or the maximum (Eq.\eqref{maxsmoothgrad}) aggregation method. As reported for an example lesion in Fig.~\ref{fig:max-vs-avg}, the saliency maps generated with the maximum presented values that were comparable to those obtained for smaller lesions while preserving the proportion of positive and negative values according to the lesion properties. In the case of the method based on SG, the results in the following sections are obtained using the max aggregation method.

\subsection*{Assessing the contributions of input MR sequences using gradient-based saliency maps}
\hypertarget{Assessing the contributions of input MR sequences using gradient-based saliency maps}{}

\begin{figure}[!ht]
    \centering
    \includegraphics[width=0.5\textwidth]{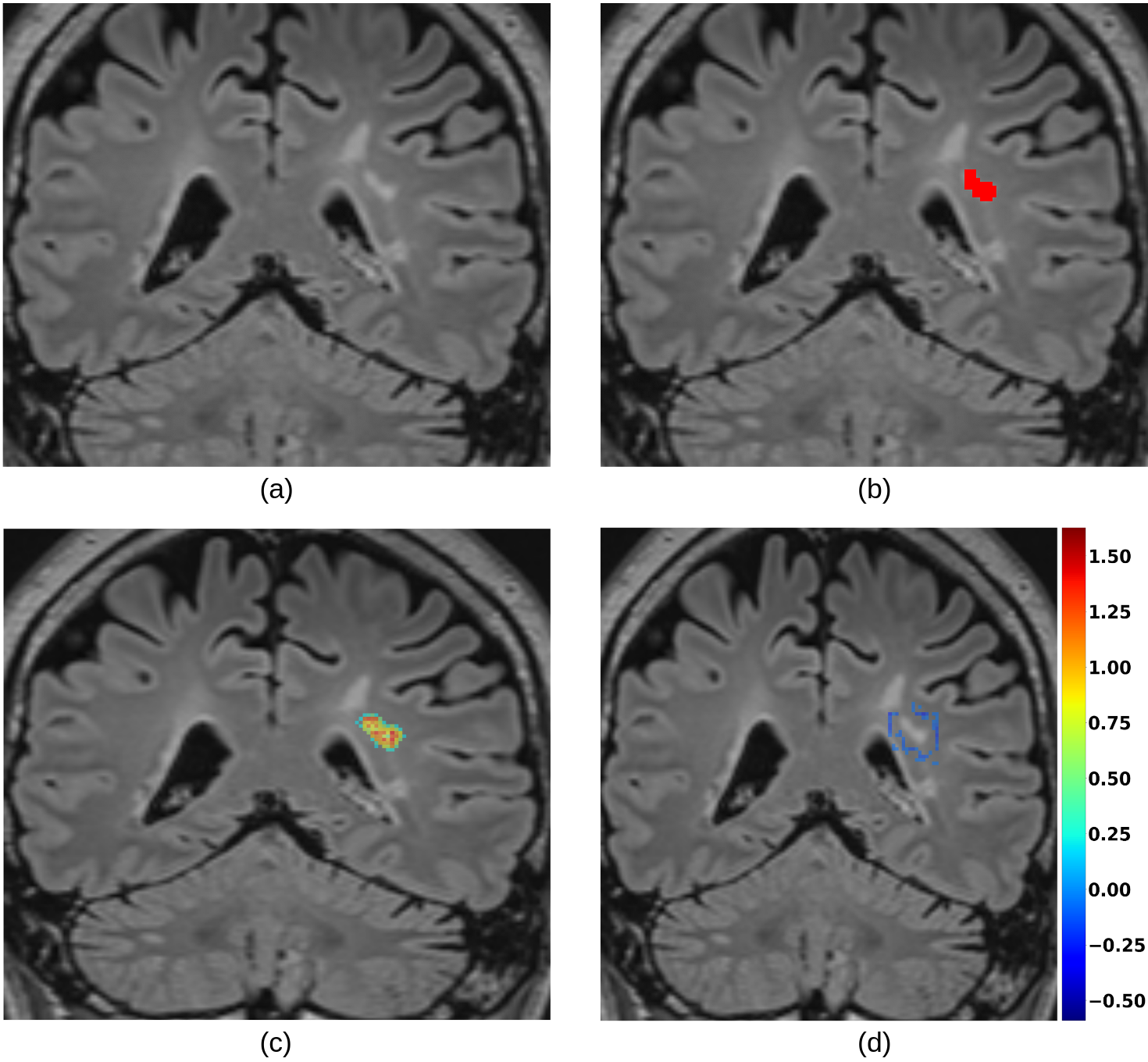}
    \caption{Example of a FLAIR image (a), the lesion domain $\Omega$ (b), and the corresponding saliency map (computed for U-Net) obtained with the proposed adaptation of SmoothGrad isolating positive (c) and negative (d) gradients. Values in [-0.05, 0.2] are not displayed to focus on most significant saliency.}
    \label{fig:location-flair}
\end{figure}

\begin{figure}[!ht]
    \centering
    \includegraphics[width=0.5\textwidth]{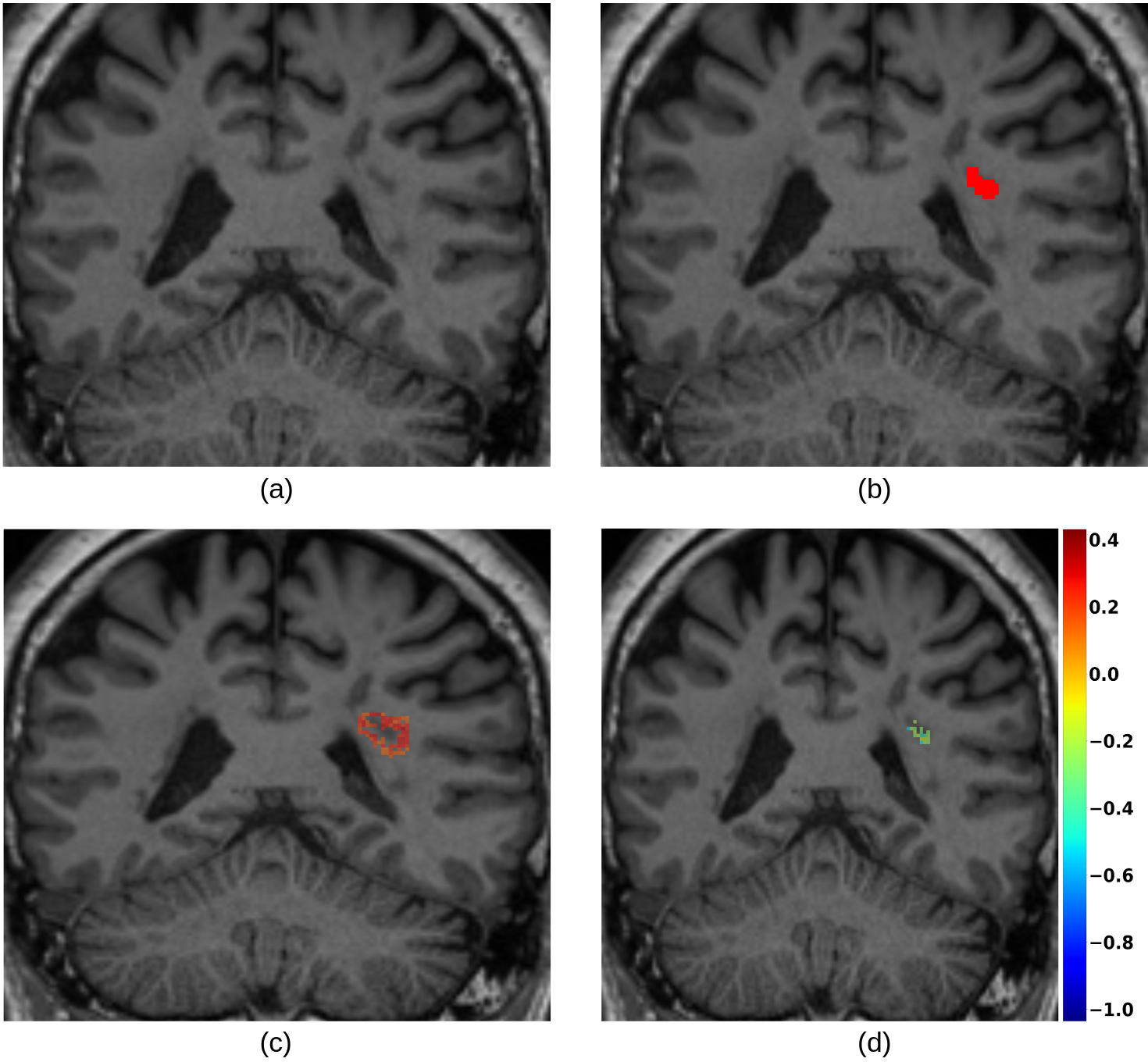}
    \caption{Example of an MPRAGE image (a), the lesion domain $\Omega$ (b), and the corresponding saliency map (computed for U-Net) obtained with the proposed adaptation of SmoothGrad isolating positive (c) and negative (d) gradients. Values in [-0.1, 0.1] are not displayed to focus on the most significant saliency.}
    \label{fig:location-mprage}
\end{figure}

\begin{figure}[!ht]
    \centering
    \includegraphics[width=0.5\textwidth]{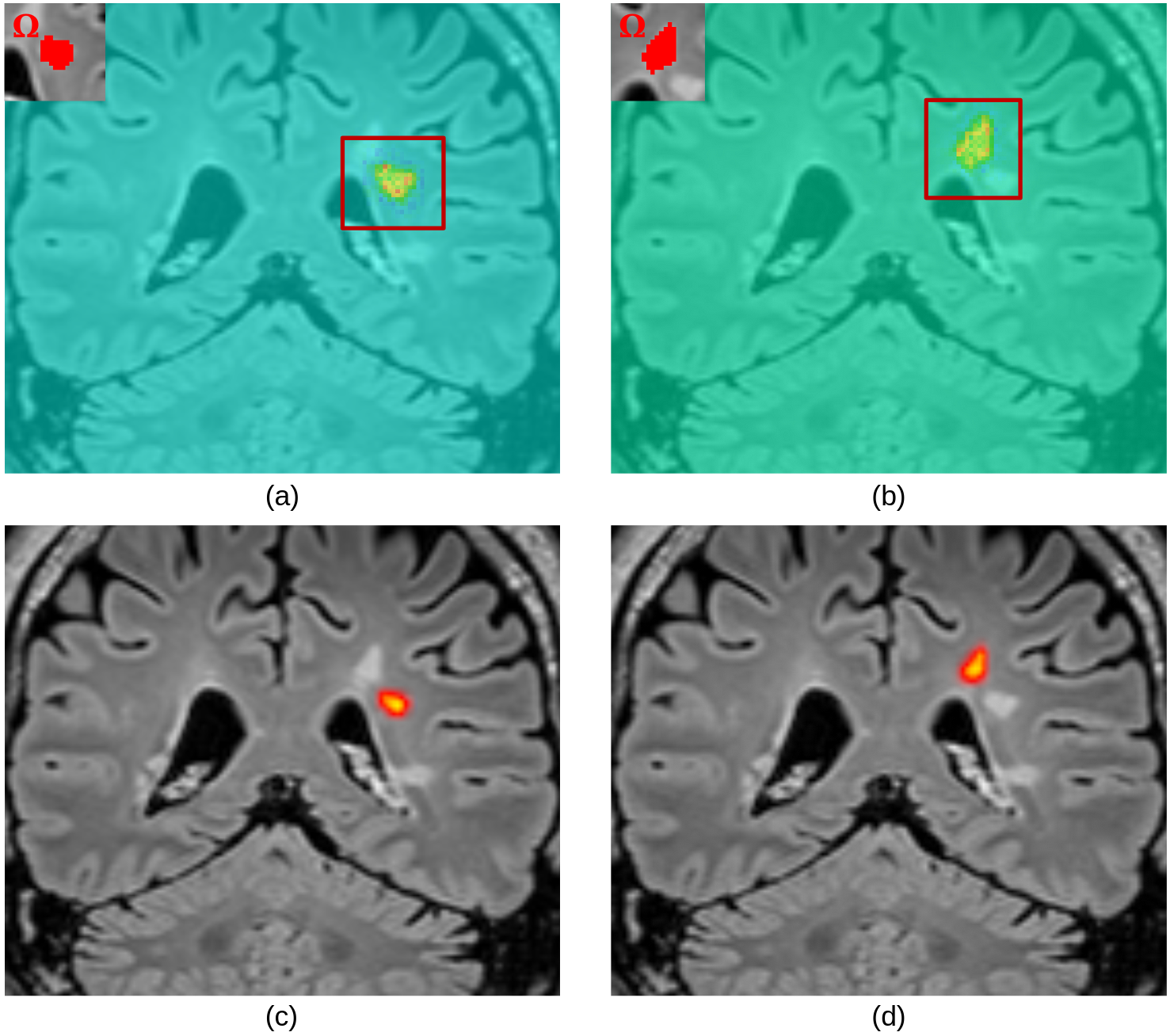}
    \caption{Saliency maps (computed for U-Net) obtained with the proposed adaptation of SmoothGrad for two close lesions (a) and (b). Heatmaps obtained with the proposed adaptation of Grad-CAM++ for the same two lesions (c) and (d). The thumbnails illustrate the lesion domains $\Omega$ used during the computation.}
    \label{fig:proximal}
\end{figure}

Saliency maps (based on SG) generated with respect to FLAIR for a true positive (TP) lesion are presented in Fig.~\ref{fig:location-flair}. Positive gradient values appeared to accumulate inside the targeted lesion domain $\Omega$ and its edges, while negative values populated its neighborhood. Following the same procedure with respect to MPRAGE, we observed an opposite trend (Fig.~\ref{fig:location-mprage}): negative values in $\Omega$ and positive around its borders. In both cases, the values dropped to zero when looking at a distance of $\sim44 mm$ from $\Omega$'s borders, and presented close-to-zero values for other lesions in the same brain regions as $\Omega$. The latter was also observed in the instance-level explanation method based on Grad-CAM++ (see Fig.~\ref{fig:proximal}).
For TP cases, we observed that positive values of gradients (median and 95\% confidence interval (CI) of 0.50140 ± 0.00072) computed with respect to FLAIR were consistently greater (in absolute value) than negative values for MPRAGE (median and 95\% CI of -0.19584 ± 0.00031). 

\subsection*{Statistical distribution of gradient values for TPs, FPs, FNs and TNs}
\hypertarget{Statistical distribution of gradient values for TPs, FPs, FNs and TNs}{}

\begin{figure}[!ht]
    \centering
    \includegraphics[width=0.7\textwidth]{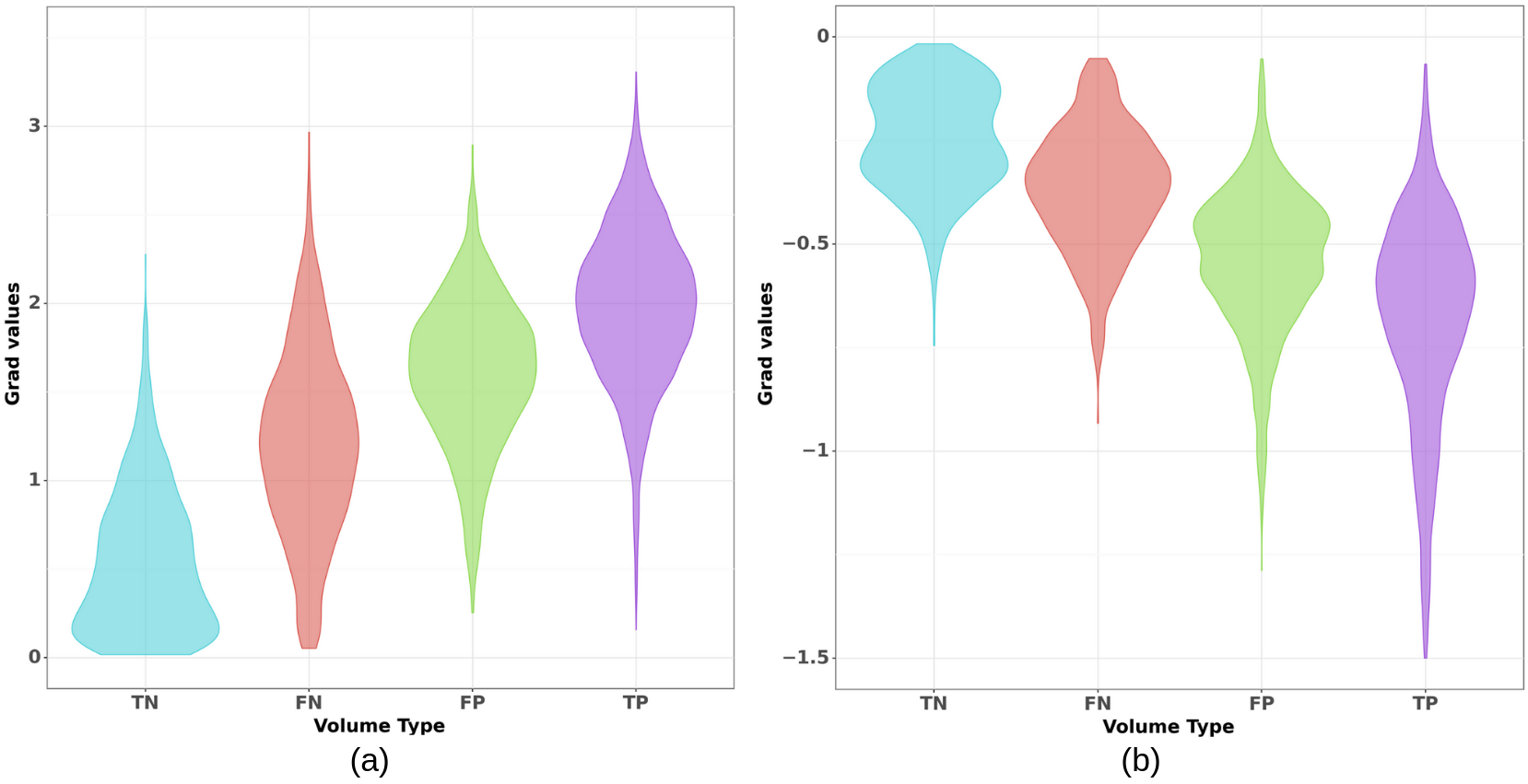}
    \caption{Violin plots representing the distribution of saliency maps (computed for U-Net) maximum (a) and minimum (b) values. The four distributions refer to TN, FN, FP and TP volumes.}
    \label{fig:violinplots}
\end{figure}

Fig.~\ref{fig:violinplots} reports the distribution of maximum (a) and minimum (b) gradients values --- with respect to FLAIR --- for TN, FN, FP and TP volumes. The median value and 95\% CI for positive values in each group of volumes were, respectively: $0.465\pm 0.025$ (TN), $1.202\pm 0.036$ (FN), $1.630\pm 0.019$ (FP), and $1.997\pm 0.016$ (TP). The median value and 95\% CI for negative values were, respectively: $-0.245\pm 0.008$ (TN), $-0.359\pm 0.010$ (FN), $-0.522\pm 0.008$ (FP), and $-0.639\pm 0.009$ (TP). The Mann Whitney U tests run on all pairs of groups reported a $p$-value < 0.001.

\subsection*{Sanity checks}
\begin{figure}[!ht]
    \centering
    \includegraphics[width=0.7\textwidth]{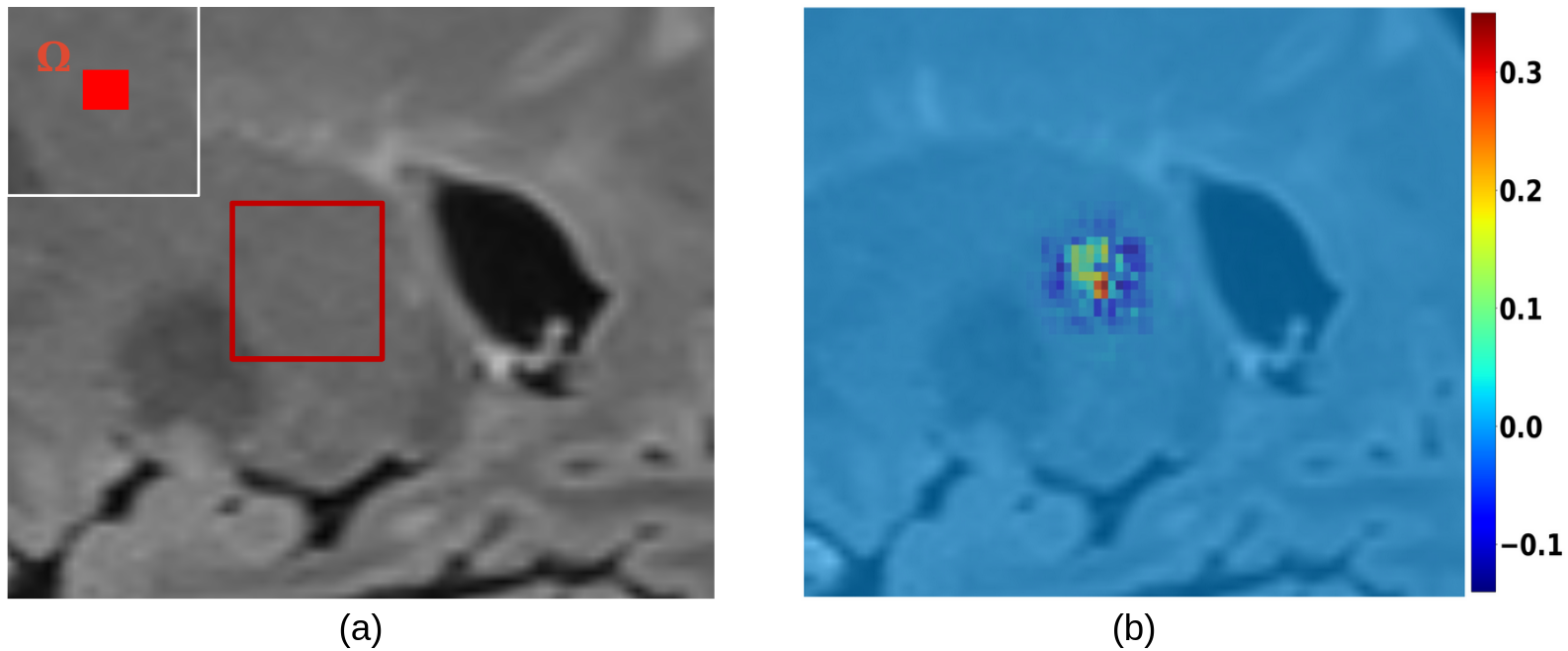}
    \caption{A region in FLAIR with healthy white matter (a) and the corresponding instance-level saliency map (computed for U-Net) obtained with the proposed adaptation of SmoothGrad (b). The color scale range is (-0.1, 0.3), well below the one in Figures \ref{fig:max-vs-avg}b and \ref{fig:max-vs-avg}d. The thumbnail illustrates the domain $\Omega$ used during the computation.}
    \label{fig:no_les}
\end{figure}

\hypertarget{Sanity checks}{}
The saliency map generated from a volume (in the WM) with no lesions resulted in gradient values in the range of TN examples (Fig.~\ref{fig:violinplots}), that is about 5 times smaller than the average one obtained in TP cases. An example of this finding is illustrated in Fig.~\ref{fig:no_les}.

\begin{figure}[!ht]
    \centering
    \includegraphics[width=0.7\textwidth]{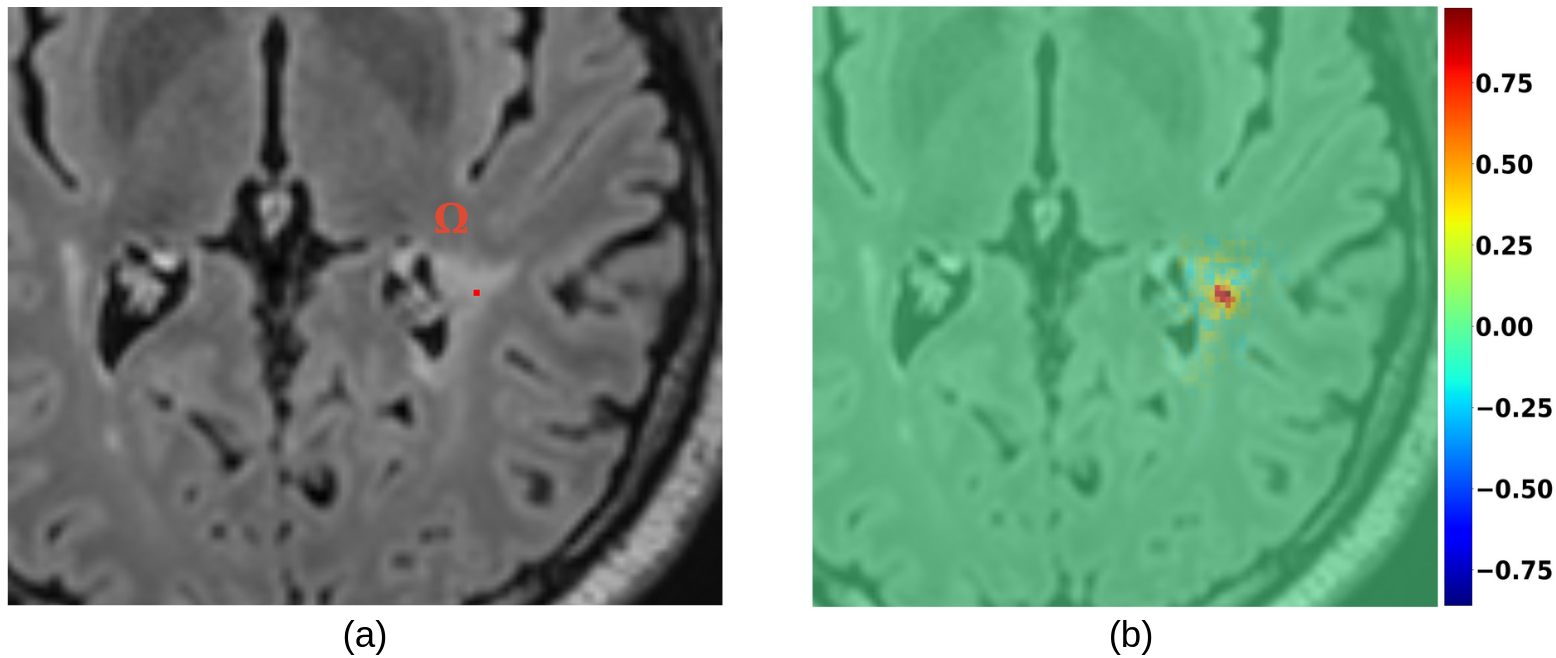}
    \caption{Axial view of an MS lesion, highlighting in red its center of mass used as domain $\Omega$ (a), and the corresponding voxel-level saliency map (computed for U-Net) obtained with the proposed adaptation of SmoothGrad (b).}
    \label{fig:single_voxel}
\end{figure}

The case of a FLAIR with a single voxel domain $\Omega$, is shown in Fig.~\ref{fig:single_voxel}. Saliency values suggest the prediction of $\Omega$ is attributed to input voxels which are in the vicinity, and within the lesion. It appears clear that lesion voxels, in the input, which are farther from $\Omega$ present a lower contribution (as it was noticed when using the average aggregation method). However, not only the input voxel corresponding to $\Omega$ has an influence.

\begin{figure}[!ht]
    \centering
    \includegraphics[width=0.5\textwidth]{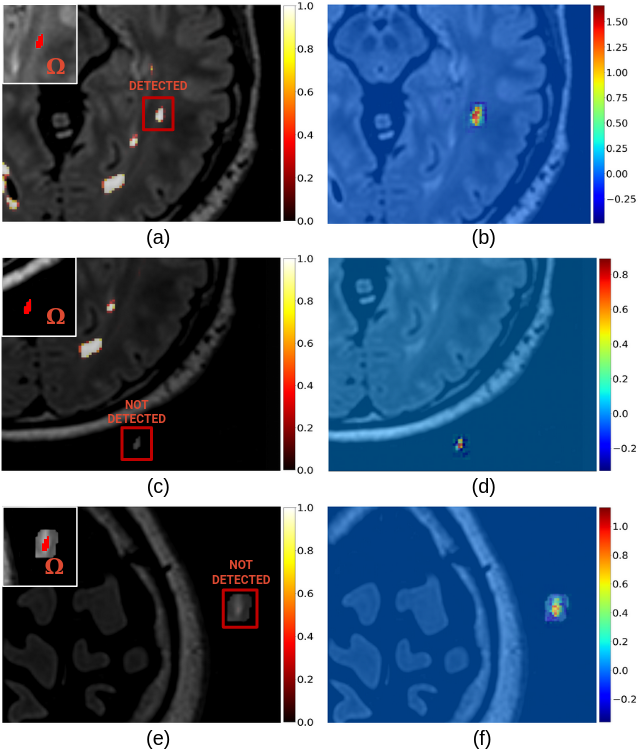}
    \caption{The prediction score before the Softmax for the case of a lesion artificially placed in the WM (a), in the background without (c) and with (e) a part of the surrounding WM. The corresponding saliency maps (b), (d) and (f), respectively. Only the synthetic lesion in the WM triggered a detection.}
    \label{fig:synthetic}
\end{figure}

When inserting a synthetic lesion in a healthy region of the WM, we obtained scores higher than 0.3, enough to trigger its segmentation. In this case, the saliency map (based on SG) resembled those obtained for true WM lesions. However, when the same lesion was placed outside the skull, the lesion domain presented scores below the threshold after the Softmax activation and, thus, the lesion was not detected. In the saliency map, we observed a few positive peak values, but the rest of the lesion volume had negative or close-to-zero gradients. Similarly to the second case, when the lesion and part of its surrounding tissue were moved outside the skull, we noticed low prediction scores. The saliency map, however, presented higher peak values than the second experiment. The prediction scores after the Softmax and the saliency maps for all these three cases were reported in Fig.~\ref{fig:synthetic}.

\subsection*{Experiment on contextual information}
\label{Experiment on contextual information}

\begin{figure}[!hb]
    \centering
    \includegraphics[width=0.7\textwidth]{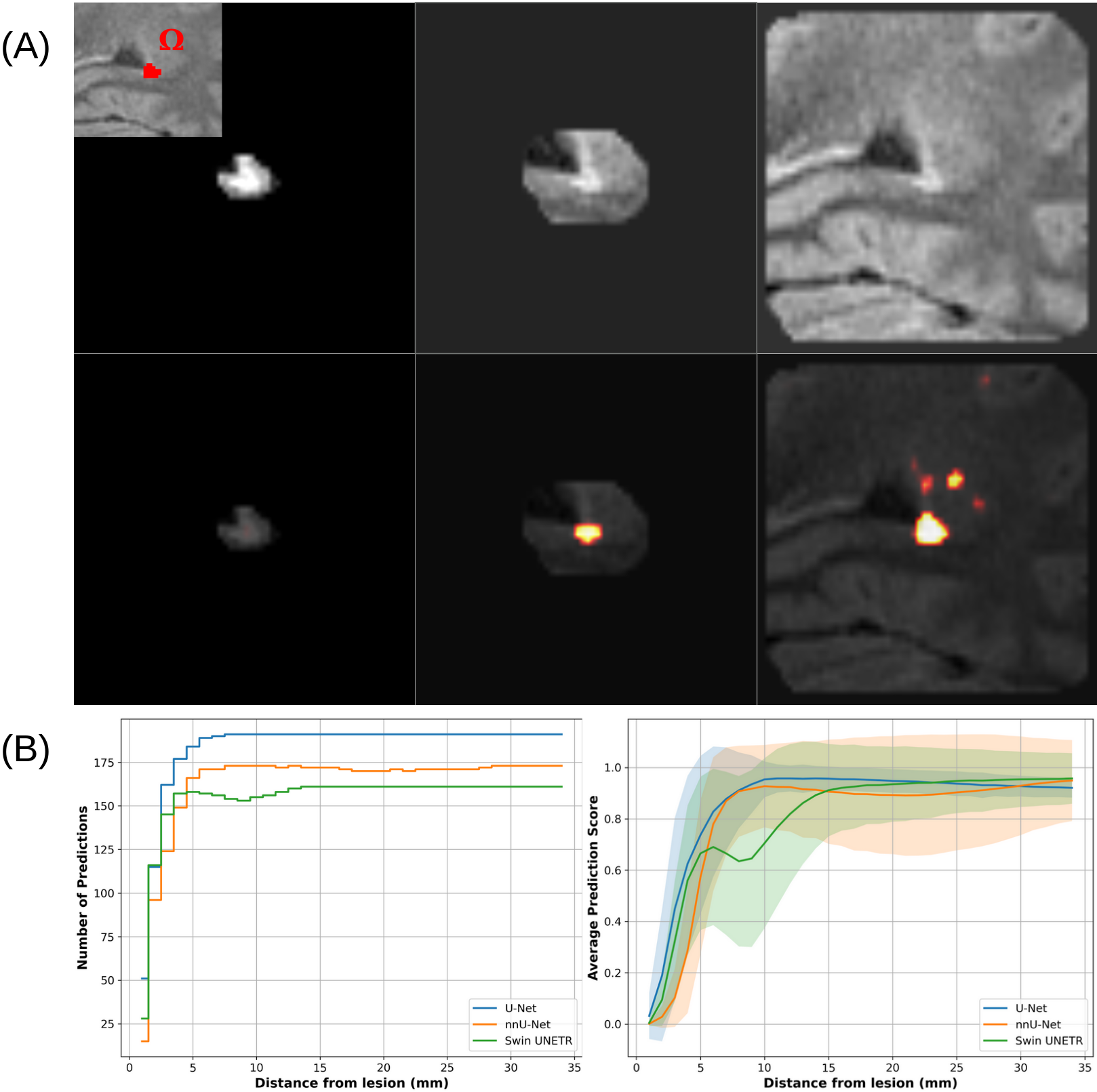}
    \caption{(A): FLAIR masked out with dilation steps 1, 5 and 24 (top), and the corresponding output probability maps computed for U-Net (bottom). (B): Plots representing the number of segmented lesions (left) and the average across patients of the mean prediction score (right) at each dilation step for three tested networks (the transparency for each line represents its standard deviation).
    \label{fig:dilation}
}
\end{figure}

The experiment on contextual information showed that, compared to a lesion without any surroundings (i.e. constant background values outside the lesion mask), the prediction score for a lesion increased when including voxels belonging to its perilesional healthy tissue. For U-Net and nnU-Net, the prediction score reached a similar plateau when adding tissue distant $12-15 mm$ from the lesion border. 
A minimum of $7 mm$ of healthy perilesional tissue allowed to correctly detect all the TP lesions, as reported in Fig.~\ref{fig:dilation}. The standard deviation reached a peak at $3 mm$ distance from lesion's border, meaning that some lesions already presented high scores while others had not yet been segmented. For Swin UNETR, the prediction score reached a plateau past $15 mm$ from the lesion border, presenting a ripple around $6 mm$. All the TP lesions were detected only when at least $13 mm$ of healthy tissue were seen by the network.

\section*{Discussion}


We introduced novel methods to generate quantitative instance-level saliency maps, pushing the explanation capabilities of basic classification and segmentation saliency methods (SG and GradCAM++) to single instance segmentation. 
In medical image analysis, the latter is crucial to reveal the information used by a model to detect and segment a specific lesion of interest among other lesions of the same class, which is a very important task in daily radiology to assess disease development and response to treatment for all multi-lesional diseases (e.g., MS, metastatic cancer).
In addition to providing explanations for a models' decision, the proposed quantitative maps can be used to optimize a network's architecture and to boost a semantic segmentation models' performance. This is shown in Spagnolo et al. \cite{spagnolo2025}, where features from saliency maps were used to increase detection F1 score and positive predictive value.

Although the test Dice scores of U-Net and nnU-Net are below that of Swin UNETR and of other state of the art methods~\cite{zhang2019,commowick2021}, there are some aspects of our approach one should consider: (i) the target metric for the loss was the normalized Dice score, which was 0.71 at test (ii) the imaging data do not contain synthetic lesions, and are not skull stripped (iii) possibly, the volume of false positive predictions is relatively small and (iv) the computational cost of saliency maps for Swin UNETR is higher. The false positive predictions of the model have been visually examined, and are mostly located in cerebral sulci and gyri. The use of brain extraction tools on our data would also have contributed to the removal of some false positives, since a few of them were hyperintensities located in non-cerebral tissue.

The distribution of values in the saliency maps generated with SG showed that FLAIR imaging had a more significant contribution to the segmentation of lesions, compared to MPRAGE. This finding allowed us to check whether the model's behavior is coherent with clinical practice. Indeed, it resonates with the fact that FLAIR offers a better contrast for common MS lesions in the WM, such as periventricular and sub-cortical, since the signal from the cerebrospinal fluid is suppressed. Only a part of these lesions are clearly visible in T1-weighted images, such as MPRAGE, which are preferred to detect cortical plaques~\cite{trip2005,nelson2008}.

For a lesion domain $\Omega$, saliency maps (based on SG) with respect to FLAIR presented positive and negative gradients, respectively distributed inside $\Omega$ and in its neighborhood (Figs. \ref{fig:location-flair} and \ref{fig:location-mprage}). Indeed, positive gradients indicate that an increase in their intensities in the MRI would suggest the presence of a lesion in $\Omega$. Conversely, voxels with negative saliency values indicate that the presence of a lesion in $\Omega$ would be suggested by a decrease in their intensities in the MRI. The antithetic distribution of saliency values in MPRAGE is due to the fact that WM lesions appear as hyperintense in FLAIR and as hypointense in MPRAGE, compared to the healthy WM.

Both of our XAI methods showed that voxels far from a lesion domain $\Omega$ did not appear to impact the prediction of voxels belonging to $\Omega$ (Fig. \ref{fig:proximal}). Since we employed a CNN, each network unit did not depend on the entire input, but on a region called receptive field. Luo et al.~\cite{Luo2016} described that the effective area of a receptive field (effective receptive field, or ERF) starts as small, and then grows during training. Furthermore, the same study described skip-connections, part of a U-Net, as a cause of the reduction of receptive field's size. In our particular case, one of the possible interpretations could be that useful features to segment a lesion were close to the lesion itself and its neighborhood, so the learned receptive field is small. For instance, it is likely that re-training our model by including the images of the contextual experiment (e.g., until dilation step 25) and labeling them as background, would lead the model to have a larger ERF. Saliency maps, after such re-training, may show a higher attention to neighbouring tissue at a greater distance from a lesion border.

Peak values of saliency maps generated for the four groups of volumes (TP, FP, FN and TN) presented distributions that are significantly different from each other (Fig. \ref{fig:violinplots}). This suggests that, even if not segmented, FN volumes captured the model's attention notably more than TN volumes during inference. A similar conclusion can be drawn for TP and FP volumes. In both cases, our quantitative saliency maps could help increase the sensitivity and specificity of the model. However, in the case of FN volumes, some external input would be needed to select the lesion domain for computation. Such input could be that of a neurologist (for brain lesions), or could be derived from other maps, perhaps based on prediction's uncertainty.

Experiments on synthetic lesions suggested that the location of a lesion in the WM was not as important as the intensity of voxels within the lesion and its neighborhood (Fig. \ref{fig:synthetic}). The described behavior is expected, since we used a patch-based network. This, along with FLAIR's importance over MPRAGE, would support the hypothesis that the model’s predictions rely predominantly on voxel intensities inside lesions in FLAIR.

However, the last experiment on contextual information revealed that high and stable prediction scores were related to the amount of contextual healthy brain tissue from the perilesional volume (Fig. \ref{fig:dilation}). For U-Net and nnU-Net, enriching the context around the lesion in the input resulted in an increase of prediction scores up to a distance of $12-15 mm$ from the lesion border. Past this distance, additional voxels no longer impact the prediction as seen by the plateau in Fig.~\ref{fig:dilation}(B). A possible conclusion is that a patch size of $96mm^{3}$ could have been unnecessarily large for most lesions. Furthermore, it would be interesting to test if the patch size during training would have a potential influence on the plateau's position of Fig.~\ref{fig:dilation}(B). We might expect that bigger patches would make the model require more contextual information to segment a lesion. The trajectory of prediction scores from Swin UNETR followed the same overall trend as the other two networks. However, certain architectural characteristics — such as the use of partitioned tiles ($8 mm$ wide) applied to input patches — may make the model more sensitive to our experimental setup.

Regarding the potential of using the proposed XAI to improve model performance, the maximum and minimum values of XAI maps in Fig.~\ref{fig:violinplots} alone did not show enough discriminatory power between the different groups. However, a simple linear model relying on saliency radiomics allowed to reduce FPs, which is demonstrated in Spagnolo et al. (2025)\cite{spagnolo2025} with a clear example of how our method can be relevant to the development and optimization of a segmentation model, and ultimately to the clinics.

\section*{Conclusion}

We proposed novel XAI methods to provide quantitative instance-level explanations for segmentation, which we applied to the specific case of MS lesion segmentation. The analysis of the explanation maps and additional tests revealed fundamental insights into the decision mechanism of a deep neural network. The explanation maps were useful to understand the importance of the perilesional volume and improve the network's classification performances. The following experiment on the contextual information exploited by the network can guide architecture choices, such as patch size. The acquired new knowledge is crucial for AI engineers and clinical researchers, which constitutes an important step in facilitating AI integration into clinical practice. The proposed methods can potentially be applied to various segmentation architectures and tasks outside the medical imaging field, such as autonomous driving and robotics. Future research could leverage uncertainty maps to generate XAI maps of false negative examples, and further boost performance. Additionally, the pruning of false predictions could be studied separately for different areas of the brain.

\section*{Data Availability}

The datasets generated during and/or analysed during the current study are available from the corresponding author on reasonable request.

\bibliography{refs}



\section*{Acknowledgements}


This work was supported by the Hasler Foundation with the project MSxplain number 21042, the Swiss National Science Foundation (SNSF) with the project 205320\_219430, and the Swiss Cancer Research foundation with the project TARGET (KFS-5549-02-2022-R). We acknowledge access to the expertise of the CIBM Center for Biomedical Imaging, a Swiss research center of excellence founded and supported by CHUV, UNIL, EPFL, UNIGE and HUG.

\section*{Author contributions statement}


F.S., V.A. and A.D. conceived the experiments, F.S. conducted the experiments, L.M-G, M.O.P. and F.S. processed and organized the data, F.S. and N.M. created the software used in the work. V.A., A.D. and C.G. supervised the work. F.S. led the writing of the manuscript together with V.A. and A.D. All authors reviewed the manuscript.

\section*{Additional information}

\subsection*{Competing interests}

The University Hospital Basel (USB) and the Research Center for Clinical neuroimmunology and Neuroscience (RC2NB), as the employers of Cristina Granziera,  have received the following fees which were used exclusively for ( research support from Siemens, GeNeuro, Genzyme-Sanofi, Biogen, Roche. They also have received  advisory board and consultancy fees from Actelion, Genzyme-Sanofi, Novartis, GeNeuro, Merck, Biogen and Roche; as well as speaker fees from Genzyme-Sanofi, Novartis, GeNeuro, Merck, Biogen and Roche.




\end{document}